\documentclass[prd,superscriptaddress,amsfonts,amssymb,amsmath,showpacs,twocolumn]{revtex4-2}
\usepackage{bm}
\usepackage{amsfonts}
\usepackage{latexsym}
\usepackage[latin1]{inputenc}
\usepackage{graphicx}
\usepackage{amsmath}
\usepackage{palatino}
\usepackage{mathpazo}
\usepackage{textcomp}
\usepackage{float}
\usepackage{booktabs}
\usepackage{dcolumn}
\usepackage{hyperref}

\hypersetup{colorlinks,citecolor=red}

\usepackage{amsmath}
\usepackage{xcolor}
\usepackage{orcidlink}
\usepackage[caption=false]{subfig}
\usepackage{commath}
\def\jnl@style{\it}
\def\aaref@jnl#1{{\jnl@style#1}}

\def\aaref@jnl#1{{\jnl@style#1}}

\def\aj{\aaref@jnl{AJ}}                   
\def\apj{\aaref@jnl{ApJ}}                 
\def\apjl{\aaref@jnl{ApJ}}                
\def\apjs{\aaref@jnl{ApJS}}               
\def\apss{\aaref@jnl{Ap\&SS}}             
\def\aap{\aaref@jnl{A\&A}}                
\def\aapr{\aaref@jnl{A\&A~Rev.}}          
\def\aaps{\aaref@jnl{A\&AS}}              
\def\mnras{\aaref@jnl{Mon.~Not.~Roy.~Astron.~Soc.}}             
\def\prd{\aaref@jnl{Phys.~Rev.~D}}        
\def\prc{\aaref@jnl{Phys.~Rev.~C}}  
\def\prl{\aaref@jnl{Phys.~Rev.~Lett.}}    
\def\qjras{\aaref@jnl{QJRAS}}             
\def\skytel{\aaref@jnl{S\&T}}             
\def\ssr{\aaref@jnl{Space~Sci.~Rev.}}     
\def\zap{\aaref@jnl{ZAp}}                 
\def\nat{\aaref@jnl{Nature}}              
\def\aplett{\aaref@jnl{Astrophys.~Lett.}} 
\def\apspr{\aaref@jnl{Astrophys.~Space~Phys.~Res.}} 
\def\physrep{\aaref@jnl{Phys.~Rep.}}      
\def\physscr{\aaref@jnl{Phys.~Scr}}       
\def\commat{\aaref@jnl{Comm.~Math.~Phys.}}              
\def\science{\aaref@jnl{Science}}               
\def\cqg{\aaref@jnl{Classical Quant.~Grav.}}            
\def\jpcs{\aaref@jnl{JPCS}}                                     
\def\ijmpd{\aaref@jnl{Int.~J.~Mod.~Phys.~D}}                    
\def\grg{\aaref@jnl{Gen.~Relat.~Gravit.}}               
\def\rpp{\aaref@jnl{Rep.~Prog.~Phys.}}          
\def\npa{\aaref@jnl{Nucl.~Phys.~A}}        
\def\lrr{\aaref@jnl{Living Rev.~Rel.}}                   
\def\jcap{\aaref@jnl{J.~Cosmology Astropart.~Phys.}}    
\def\rmp{\aaref@jnl{Rev.~Mod.~Phys.}}   
\def\epjc{\aaref@jnl{Eur.~Phys.~J.~C}}

\allowdisplaybreaks[1]

\begin{document}

\color{black}

\title{Thin-Shell Gravastar Model in $f(Q,T)$ Gravity}

\author{Sneha Pradhan\orcidlink{0000-0002-3223-4085}}
\email{snehapradhan2211@gmail.com}
\affiliation{Department of Mathematics, Birla Institute of Technology and
Science-Pilani,\\ Hyderabad Campus, Hyderabad-500078, India.}

\author{Debasmita Mohanty\orcidlink{0009-0006-8118-5327}}
\email{newdebasmita@gmail.com}
\affiliation{Department of Mathematics, Birla Institute of Technology and
Science-Pilani,\\ Hyderabad Campus, Hyderabad-500078, India.}

\author{P.K. Sahoo\orcidlink{0000-0003-2130-8832}}
\email{pksahoo@hyderabad.bits-pilani.ac.in}
\affiliation{Department of Mathematics, Birla Institute of Technology and
Science-Pilani,\\ Hyderabad Campus, Hyderabad-500078, India.}
\affiliation{Faculty of Mathematics \& Computer Science, Transilvania University of Brasov, Eroilor 29, Brasov, Romania}
\date{\today}

\begin{abstract}

In the last few decades, gravastars have been proposed as an alternative to black holes. The stability of the gravastar has been studied in many modified theories of gravity along with Einstein's GR. The $f(Q, T)$ gravity, a successfully modified theory of gravity for describing the current accelerated expansion of the Universe, has been used in this article to study  gravastar in different aspects.  According to Mazur and Mottola \cite{mazur,mottola}, it has three regions with three different equations of state. Here in this work, we have studied the interior of the gravastar by considering the $p=-\rho$ EoS to describe the dark sector for the interior region. The next region is a thin shell of ultrarelativistic stiff fluid, in which we have investigated several physical properties, viz., the proper length, energy, entropy, surface energy density, etc. In addition, we have studied the surface redshift and speed of sound to check the potential stability of our proposed thin-shell gravastar model. Apart from that, we have used the entropy maximization technique to verify the stability of the gravastar model. The gravastar's outer region is a complete vacuum described by exterior Schwarzschild geometry. Finally, we have presented a stable gravastar model which is singularity-free and devoid of any incompleteness in classical black hole theory. \\


\end{abstract}

\maketitle

\section{Introduction}\label{sec:1}

There has been a large scientific interest in understanding the problems in both cosmology and astrophysics during the past few decades. Compact objects are a crucial source for this reason because they provide a platform to test many pertinent ideas in the high-density domain. The Gravitationally Vacuum Condense Star, or simply gravastar, is an excellent notion for an extremely compact object that addresses the singularity problems in classical black hole (CBH) theory. It was first postulated by Mazur and Mottola \cite{mazur,mottola}.  They construct a cold, compact object with an internal de Sitter condensate phase and an exterior Schwarzschild geometry of any total mass M that is free of all known limitations on the known CBH. As a result, this hypothesis has gained popularity among researchers and it could be seen as an alternative for the CBH.\\

 The gravastar, in particular, has three separate zones with different equations of states (EoS), according to Mazur and Mottola's model:  \begin{enumerate}
     \item An internal region that is full of dark energy with an isotropic de Sitter vacuum situation.
 \item  An intermediate thin shell consists of stiff fluid matter. 
 \item The outer area is completely vacuum, and Schwarzschild geometry represents this situation appropriately.
 \end{enumerate} 
 
 Recent studies on the brightness of type Ia distant supernovae \cite{AR,SJ,NA} indicate that the universe is expanding more quickly than previously thought, which suggests that the universe's pressure $p$ and energy density $\rho$ should contradict the strong energy condition, that is, $\rho+3 p<0$. "Dark energy" is the substance that causes this requirement to be fulfilled at some point in the evolution of the universe \cite{VS,PJE,TP}. There are several substances for the status of dark energy.  The most well-known contender is a non-vanishing cosmological constant, which is equivalent to the fluid that satisfies the EoS $p=-\rho$. There are two interfaces (junctions) located at $R_1$ and $R_2$ apart from the center, where $R_1$ and $R_2$ stand for the thin shell's interior and outer radii. The presence of stiff matter on the shell with thickness $R_2-R_1=\epsilon <<1$ is required to provide the system's stability, which is achieved by exerting an inward force to counteract the repulsion from within.\\

 Astrophysicists proposed a new solution of a compact, spherically symmetric astrophysical phenomenon, known as a gravastar, to solve the singularity problem in black hole geometry. There are several arguments for and against the theory that gravitational waves (GW), detected by LIGO, are the consequence of merging gravastars or black holes, despite the fact that no experimental observations or discoveries of gravastars have yet been made. A method for identifying gravastar was devised by Sakai et al. \cite{sakai} by looking at the gravastar shadows. Since black holes don't exhibit microlensing effects of maximal brightness, Kubo and Sakai \cite{kubo}  hypothesized that gravitational lensing may be used to find gravastars. The finding of GW150914 \cite{cardoso, Cardoso} by interferometric LIGO detectors increased the likelihood that ringdown signals originated from sources without an event horizon. In a recent examination of the picture taken by the First M87 Event Horizon Telescope (EHT), a shadow that resembled a gravastar has been discovered \cite{akiyama}.\\
  One could observe that there are numerous publications on the gravastar available in the literature that focuses on various mathematical and physical problems in the framework of general relativity postulated by Albert Einstein \cite{visser,carter,bilic,lobo,Lobo,cattoen,ghosh,Ghosh,rahaman,usmani}. Bilic et al. \cite{bilic} replaced the de Sitter interior with a Chaplygin gas equation of state and saw the system as a Born-Infield phantom gravastar to examine the gravastar's interior, whereas Lobo \cite{lobo} replaces the inner vacuum with dark energy. Although it is commonly known that Einstein's general relativity is an exceptional tool for revealing many hidden mysteries of nature, certain observable evidence of the expanding universe and the existence of dark matter has put a theoretical challenge to this theory. Hence, a number of modified theories have been put over time, like $ f(R), f(Q), f(T), f(R,T), f(Q,T)$ gravity, etc. The $f(R), f(R,T)$ gravity which is based upon the Riemannian geometry in which Ricci scalar curvature plays an important role. Another way to represent the gravitational interaction between two particles in space-time is by torsion and non-metricity upon which $f(T)$ and $f(Q)$ gravity theory has been built, respectively. In the current project, our objective is to investigate the gravastar using one of the alternative theories of gravity, $f(Q,T)$ gravity, and to examine many physical characteristics and stability of the object. The $f(Q,T)$ gravity is the extension of the symmetric teleparallel gravity in which the gravitational action is determined by any function $f$ of the nonmetricity $Q$ and the trace of the matter energy-momentum tensor $T$, such that $L= f(Q,T)$.  There are very few articles in which compact object has been studied under the framework of $f(Q,T)$ gravity \cite{Tayde}. Xu et al. have investigated the cosmological implication of this theory, and they have obtained the cosmological evolution equation for isotropy, homogeneous, flat geometry \cite{xu}. In \cite{godani}, the author has investigated the different  FRW models with three specific forms of $f(Q,T)$ gravity models. One could see the references to the recent work of gravastar in the framework of modified gravity \cite{sharif,S1,S2,das,sahoo,bhatti}. In the article, \cite{pradhan} researchers have studied the gravastar model in $f(Q)$ gravity. Ghosh et al. \cite{amit} has studied the gravastar in Rastall gravity. In the work \cite{NG1}, the author has studied traversable wormhole solutions in the presence of the scalar field. Wormhole solutions in $f(R, T)$ gravity have been studied  in \cite{NG2}. Elizalde et al. \cite{Elizalde} discussed the cosmological dynamics in $R^2$  gravity  with logarithmic trace term. Godani and Samanta \cite{Godani & Gauranga} discussed the gravitational lensing effect in traversable wormholes. In \cite{Godani & Samanta}, the researchers have investigated wormhole solutions with scalar field and electric charge in modified gravity.  In \cite{samanta}, the authors studied the cosmologically stable $f(R)$ model and wormhole solutions. Salvatore et al. \cite{capozziello} studied the non-local gravity wormholes, and they obtained  stable and traversable wormhole solutions. Shamir et al. \cite{A1} has explored the behavior of anisotropic compact stars in $f(R,\phi)$ gravity. The Bardeen compact stars in Modified $f(R)$ gravity have been researched in the work \cite{A2}. 
 \\

Our paper is organized as follows: In sec \ref{sec:1} we have given a brief introduction to the gravastar model and the recent research work regarding that. After that in sec \ref{sec:2} we provide the geometrical aspects of $f(Q,T)$ gravity.  In sec \ref{sec:3} we have derived the modified field equation and the modified energy conservation equation in $f(Q,T)$ gravity.  Sec \ref{sec:4} gives the solution of the field equation for different regions using different EoS. After that in sec \ref{sec:5} we have studied the junction requirement and EoS and we have obtained the limiting range for the radius of the gravastar. The physical features of the model has been analyzed in sec \ref{sec:6}. The most important thing is to check the stability of the model which is given in sec \ref{sec:7}. Finally, we provide the conclusion of our analysis in sec \ref{sec:8}.

\section{Construction of $f(Q,T)$Gravity}\label{sec:2} 
The $f(Q,T)$ theory of gravity which introduces an arbitrary function of scalar non-metricity $Q$ and trace $T$ of the matter energy-momentum tensor, is an intriguing modification to Einstein's theory of gravity.
The action of $f(Q,T)$ theory coupled with matter Lagrangian $\mathcal{L}_m$ is given by \cite{Xu}
   
\begin{equation}
\label{1}
S=\int \sqrt{-g}\left[\frac{1}{16 \pi }f(Q,T)+\mathcal{L}_m\right] d^4x,
\end{equation}
where $g$ represents the determinant of $g_{\mu \nu}$. The non-metricity and disformation tensor is defined as
\begin{eqnarray}
\label{2}
Q\equiv- g^{\mu\nu}\left(L^\alpha_{\,\,\beta\mu}L^\beta_{\,\,\nu\alpha}-L^\alpha_{\,\,\beta\alpha}L^\beta_{\,\,\mu\nu}\right),\\
L^\lambda_{\,\,\,\,\mu\nu}=-\frac{1}{2}g^{\lambda\gamma}\left(\nabla_{\nu}g_{\mu\gamma}+\nabla_{\mu}g_{\gamma\nu}-\nabla_{\gamma}g_{\mu\nu}\right).
\end{eqnarray}

The non-metricity tensor is defined as the covariant derivative of the metric tensor, and its explicit form is
\begin{equation}
\label{4}
Q_{\alpha\mu\nu}\equiv \nabla_\alpha g_{\mu\nu}.
\end{equation}
with the trace of a non-metricity tensor as
\begin{equation*}
Q_{\lambda}=Q_{\lambda\,\,\,\,\,\mu}^{\,\,\,\mu}, \quad \quad \tilde{Q}_{\lambda}=Q^{\mu}_{\,\,\,\,\lambda\mu} .
\end{equation*}

The Superpotential $P_{\,\,\mu\nu}^{\lambda}$ is defined as
\begin{equation}
\label{5}
P_{\,\,\,\,\mu\nu}^{\lambda}=-\frac{1}{2}L^\lambda_{\,\,\,\,\mu\nu}+\frac{1}{4}\left(Q^{\lambda}-\tilde{Q^{\lambda}}\right)g_{\mu\nu}-\frac{1}{4}\delta^{\lambda}_{\,\,(\mu\,} Q_{\nu)},
\end{equation}
giving the relation of scalar non-metricity as
\begin{equation}
\label{6}
Q=-Q_{\lambda\mu\nu}P^{\lambda\mu\nu}.
\end{equation}
 The field equations of $f(Q, T)$ theory by varying the action \eqref{1} with respect to the metric tensor inverse $g^{\mu\nu}$ is obtained as
\begin{multline}
\label{7}
-\frac{2}{\sqrt{-g}}\nabla_{\lambda}\left(f_{Q}\sqrt{-g}\,P^{\lambda}_{\,\,\,\,\mu\nu}\right)-\frac{1}{2}f\,g_{\mu\nu}+f_{T}\left(T_{\mu\nu}+\Theta_{\mu\nu}\right)\\
-f_{Q}\left(P_{\mu\lambda\alpha}Q_{\nu}^{\,\,\,\lambda\alpha}-2Q^{\lambda\alpha}_{\,\,\,\,\,\,\,\,\mu}\, P_{\lambda\alpha\nu}\right)=8\pi T_{\mu\nu}.
\end{multline}
The terms used in the above are defined as
\begin{eqnarray}
\label{8}
\Theta_{\mu\nu} &=& g^{\alpha\beta}\frac{\delta T_{\alpha\beta}}{\delta g^{\mu\nu}}, \quad  T_{\mu\nu}=-\frac{2}{\sqrt{-g}}\frac{\delta(\sqrt{-g}\mathcal{L}_m)}{\delta g^{\mu\nu}},\\
\label{9}
f_T &=& \frac{\partial f(Q,T)}{\partial T}, \quad
f_Q=\frac{\partial f(Q,T)}{\partial Q}.
\end{eqnarray}
Where $T_{\mu \nu}$ is known as the energy-momentum tensor.

\section{Modified Field Equation in $f(Q,T)$ }\label{sec:3}
To derive the modified field equation let's take the static, spherically symmetric line element given by,
\begin{equation}
    ds^2 =e^{\nu} dt^2 -e^{\lambda} dr^2 -r^2 (d\theta^2 +sin^2 \theta d\phi^2).
\end{equation}
To describe the fluid distribution we are going to take the energy-momentum tensor in the form :
\begin{equation}
    T_{\mu \nu} = (\rho+p_t) u_{\mu} u_{\nu} - p_t \delta_{\mu \nu} + (p_r -p_t) v_{\mu} v_{\nu},
\end{equation}
where $\rho$ is the density of the fluid, $p_r$ and $p_t$ are the pressures of the fluid in the direction of $u_{\mu}$(radial pressure) and orthogonal to $u_{\nu}$ (tangential pressure) respectively. $u_{\mu}$ is the time like four-velocity vector. $v_{\mu}$ is the unit space-like vector in the direction of the radial coordinate. Therefore the stress energy momentum tensor $T_{\mu \nu}$ and the components of $\Theta_{\mu\nu}$ can be expressed as,
\begin{equation}
\begin{gathered}
    T_{\mu\nu}=diag(e^{\nu}\rho, e^{\lambda}p_r,r^{2} p_t, r^2 p_t sin^{2}\theta)\\
    \Theta_{11}=-e^{\nu}(P+2\rho), \, \Theta_{22}= e^{\lambda}(P-2p_r),\\
    \Theta_{33}=r^2(P-2p_t), \,\, \Theta_{44}=r^2 sin^2\theta  (P-2p_t) .
\end{gathered}
\end{equation}
Where we have taken the Lagrangian matter density $\mathcal{L}_\mathrm{m}=-P=- \frac{p_r +2 p_t}{3}$.
By utilizing  the aforementioned constraints the derived modified field equation for spherically symmetric metric in $f(Q,T)$ gravity is,

\begin{widetext}
    \begin{equation}
        8\pi\rho = \frac{1}{2r^2 e^{\lambda}}  [2rf_{Q Q} Q'(e^{\lambda} -1)+ f_{Q}[(e^{\lambda} -1)(2+r\nu')\\+(1+e^{\lambda})r \lambda']+fr^{2}e^{\lambda}]-f_T [P+\rho],
    \end{equation}
\begin{equation}
     8\pi p_r = - \frac{1}{2r^2 e^{\lambda}}  [2rf_{Q Q} Q'(e^{\lambda} -1)+ f_{Q}[(e^{\lambda} -1)\\(2+r\nu'+r\lambda')-2r\nu']+fr^{2}e^{\lambda}]+f_T [P-p_r],
\end{equation}
\begin{equation}
     8\pi p_t= - \frac{1}{4re^{\lambda}}  [-2r f_{Q Q} Q' \nu' + f_{Q}[2\nu'(e^{\lambda} -2)- r\nu'^{2}+\lambda'(2e^{\lambda}+r\nu')-2r\nu'']+2f re^{\lambda}]
     +f_T [P-p_t] .
\end{equation}
\end{widetext}

Now here we are going to take a particular functional form of $f(Q,T)$ gravity as $f(Q,T)=\alpha \,Q+\beta \,T$. One can see there are many references \cite{Xu,Tayde,xu} in which this cosmological models has been studied widely. Then we can rewrite the field equation in a as like :

\begin{equation}
   e^{-\lambda} \left(\frac{\lambda'}{r}-\frac{1}{r^2}\right)+\frac{1}{r^2} =\rho^{eff},
   \label{eq:16}
\end{equation}
\begin{equation}
    e^{-\lambda} \left(\frac{\nu'}{r}+\frac{1}{r^2}\right)-\frac{1}{r^2}=p_r^{eff},
    \label{eq:17}
\end{equation}
\begin{equation}
    e^{-\lambda}\left(\frac{\nu"}{2}-\frac{\lambda' \nu'}{4}+\frac{\nu'^2}{4}+\frac{\nu'-\lambda'}{2r}\right)=p_t^{eff},
    \label{eq:18}
\end{equation}
Where,
\begin{equation}\label{eq:19}
     \rho^{eff}=  \frac{8 \pi \rho}{\alpha}+\frac{\beta}{3\alpha}(3 \rho+p_r +2 p_t)-\frac{\beta}{2\alpha}(\rho-p_r-2p_t),
\end{equation}
 \begin{equation}\label{eq:20}
     p_r^{eff}=\frac{8 \pi p_r}{\alpha}-\frac{2 \beta}{3 \alpha}(p_t-p_r)+\frac{\beta}{2 \alpha}(\rho-p_r-2p_t),
 \end{equation} 
 \begin{equation}\label{eq:21}
      p_t^{eff}=\frac{8 \pi p_t}{\alpha}-\frac{\beta}{3 \alpha}(p_r-p_t)+\frac{\beta}{2\alpha}(\rho-p_r-2p_t).
 \end{equation}

One can verify that for $\alpha=1, \beta=0$ i.e. for $f=Q$ the above field equation reduces to Einstein's GR. However in this article, we limit ourselves to the isotropic scenario in order to establish the simplest possibility where $p_r=p_t$. Now the energy conservation equation is given by,
\begin{equation}
    \frac{dp^{eff}}{dr}+\frac{\nu'}{2}(p^{eff}+\rho^{eff})=0.
    \label{eq:22}
\end{equation}
By using the equation (\ref{eq:19}) and (\ref{eq:20})  we get the modified energy conservation equation in $f(Q,T)$ gravity as:
\begin{equation}\label{eq:23}
     \frac{dp}{dr}+\frac{\nu'}{2}\left[(1+\frac{\beta}{8 \pi})(p+\rho)\right]+\frac{\beta}{16 \pi}(\rho'-3 p')=0.
\end{equation}
The above equation is different from that obtained in GR and can be retrieved in the limit $\beta\to 0$.
\section{Geometry of Gravastar}\label{sec:4}
We are specifically interested in the geometrical interpretation and their related analytical solution in the three different zones for the gravastar under study. It is simple to imagine the idea that the inside of the star is considered to be encircled by a thin shell made of ultrarelativistic stiff fluid, but the outside area is completely vacuum. Schwarzschild's measure is therefore assumed to be appropriate for this outer area. The shell's structure is believed to be extremely thin, with a limited width ranging $R_1=R \leq r \leq R +\epsilon=R_2$, where $r$ is the radial coordinate and $R_1, R_2$ denotes the inner and outer radius of the shell.

\subsection{Interior Region}
In the primary model proposed by Mazur and Mottola\cite{mazur,mottola} the three different zones obey the standard cosmological EoS $p=\omega \rho$, where $\omega$ is the EoS parameter takes different value for different region. Here, we suppose that an enigmatic gravitational source is present in the interior area.  Dark matter and dark energy are typically assumed to be separate entities, although there is the possibility that they are both just various representations of the same thing. For describing the dark sector in the interior region we are interested to consider the EoS is given by,
\begin{equation}
    p=-\rho.
    \label{eq:24}
\end{equation}

 For the aforementioned EoS obtaining constant critical density $\rho_c$ from the energy conservation Eq.(\ref{eq:23}), we get  the pressure for the interior region as,

\begin{equation}\label{eq:25}
    p = -\rho_c.
\end{equation}
Using Eq.(\ref{eq:25}) in field equations (\ref{eq:16}) and (\ref{eq:19}) we obtained the final expression for metric potential $e^{-\lambda(r)}$ as,

\begin{equation}
   e^{-\lambda(r)}= \frac{2 (\beta -4 \pi ) \rho_c r^3-3 c_1}{3 \alpha  r}+1.
\end{equation}
To make our solution regular at center we set the integrating constant $c_1=0$. Thus we have,
\begin{equation}\label{eq:27}
    e^{-\lambda(r)}= \frac{2 (\beta -4 \pi ) r^2  \rho_c}{3 \alpha }+1.
\end{equation}
Again using (\ref{eq:27}) we get another metric potential from (\ref{eq:17},\ref{eq:20}) as,

\begin{equation}\label{eq:28}
    e^{\nu(r)}=C_1 \left[2 (4 \pi -\beta ) \rho_c r^2-3 \alpha\right].
\end{equation}

\begin{figure}
\includegraphics[scale=0.43]{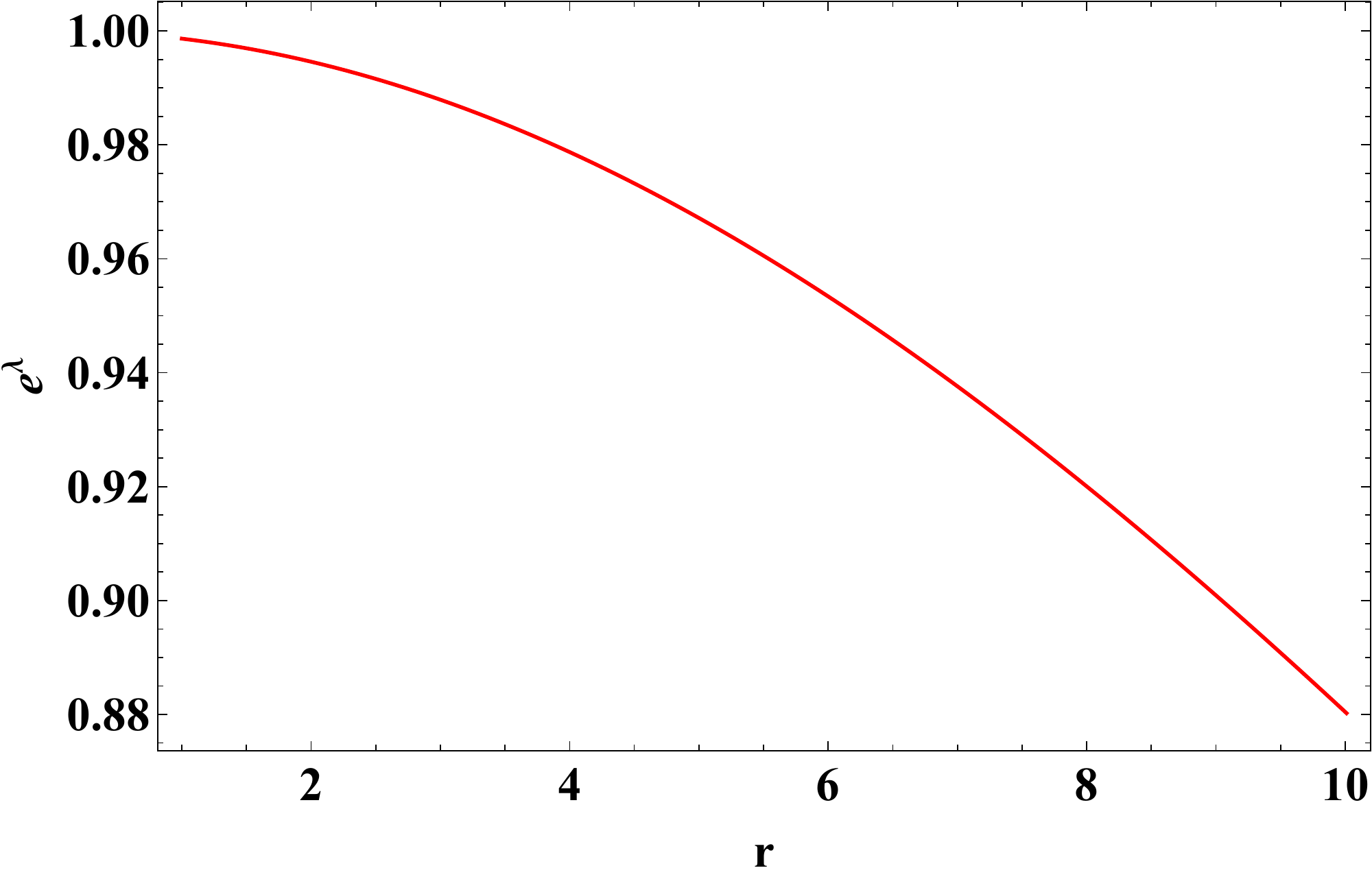}
    \caption{ Variation of the metric potential ($e^{\lambda}$)  with regard to the radial parameter $r$  for $\alpha=-4.5$, $\beta=3.4$, $\rho_c=0.001$ .}
    \label{fig:1}
\end{figure}

It is clear from the aforementioned results that there is no singularity in the inner solutions, which overcomes the issue of a classical black hole's central singularity. For more clearance, we have plotted the variation of the metric potential $e^{\lambda}$ with respect to the radial parameter $r$ in Fig.(\ref{fig:1}). 

One could physically extrapolate from the figure that there is no central singularity, along with that the metric potential is regular at $r = 0$, and it is finite and positive across the whole interior area.  
Additionally, The following equation could be used to determine the active gravitational mass of the internal region: 

\begin{equation}
\mathcal{M(R)}=\int_0^{\mathcal{R}} 4 \pi r^2 \rho dr =\frac{4}{3} \pi \mathcal{R}^3 \rho_c .
\end{equation}
Where $\mathcal{R}$ represents the radius for the interior area and $\rho_c$ is the critical density.

\subsection{Shell}
The shell is made of ultra-relativistic stiff matter and abides by the EoS $p=\rho$. Zel'dovich \cite{Zel, zel} was the pioneer of the concept of this extremely relativistic fluid known as the stiff fluid in correspondence to the cold baryonic universe. We can claim that in the current situation, this could result from thermal excitation with a very low chemical potential or from the preserved number density of the gravitational quanta at absolute zero. This kind of fluid has been widely explored by several researchers to investigate different cosmological \cite{Ms,e2,e3} and astrophysical \cite{e4,e5,e6} aspects. One may note that it is extremely challenging to solve the field equations in the non-vacuum area or the shell. However, an analytical solution could be found within the specifications of the thin shell limit, i.e. $0<e^{-\lambda(r)}<<1$. We can argue that the interior area between the two space-times must be a thin shell, as suggested by Israel \cite{Israel}. Moreover, in general, any parameter that is a function of $r$ could be considered $<< 1$ as $r\to 0$. By considering this type of approximation our field Eq.(\ref{eq:16})-Eq.(\ref{eq:18}) along with the Eq.(\ref{eq:19})-Eq.(\ref{eq:21}) reduces to:

\begin{equation}\label{eq:30}
    \alpha \left(\frac{e^{-\lambda} \lambda'(r)}{r}+\frac{1}{r^2}\right)=8 \pi \rho+ \frac{\beta}{2}(5 p+\rho),
\end{equation}
\begin{equation}\label{eq:31}
    \alpha\left(\frac{-1}{r^2}\right)=8 \pi p+\frac{\beta}{2}(\rho-3 p),
\end{equation}
\begin{equation}\label{eq:32}
    \alpha\left(\frac{-\lambda' \nu'e^{-\lambda(r) }}{4}-\frac{e^{-\lambda}\lambda'}{2r}\right)=8 \pi p+\frac{\beta}{2}(\rho-3 p).
\end{equation}

\begin{figure}
\includegraphics[scale=0.47]{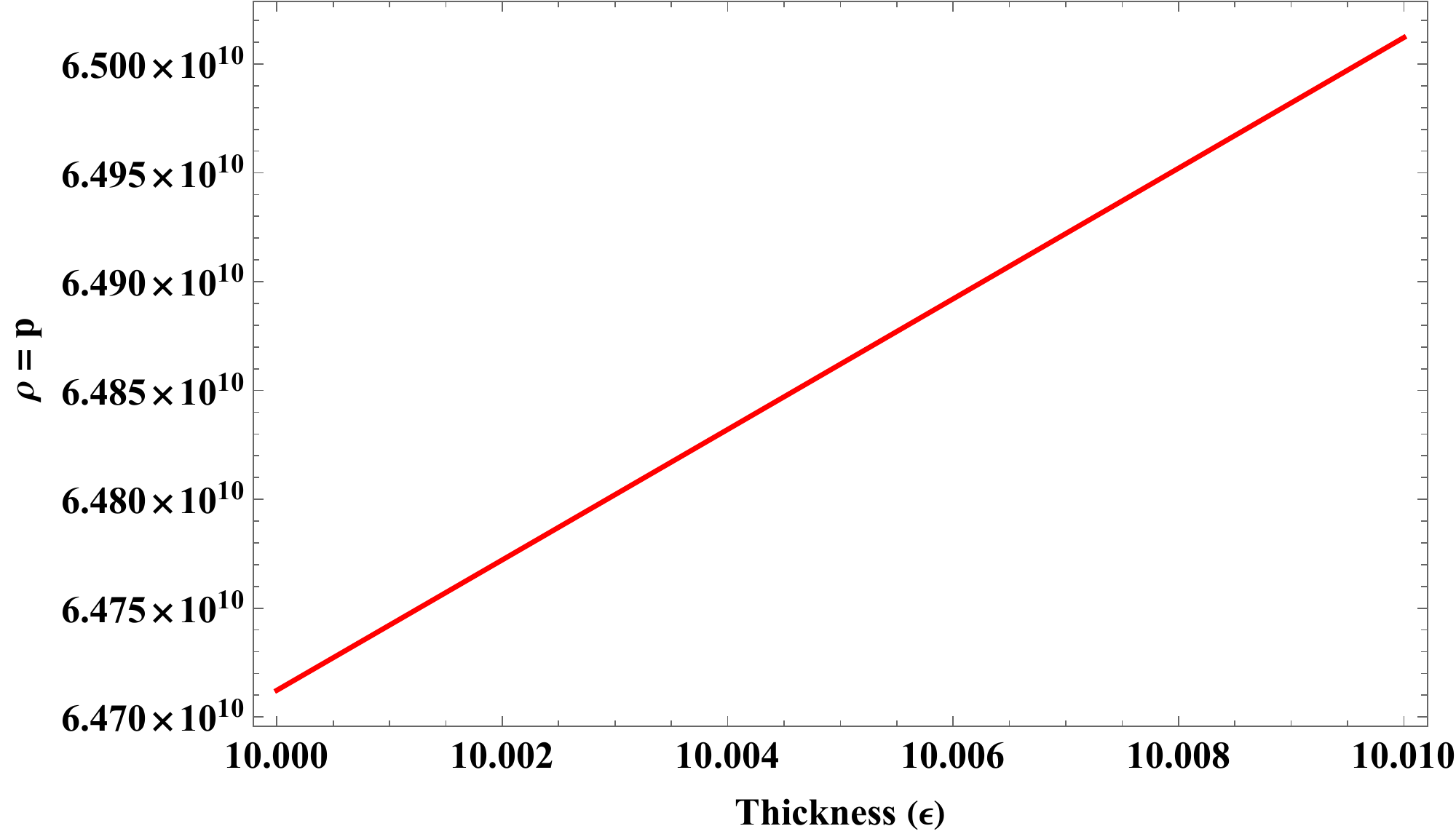}
    \caption{ Variation of the pressure or matter density $p=\rho$ \textbf{($\text{km}^{-2}$)}  with regard to the thickness $\epsilon(km)$ of the shell for $\beta=3.4$, $\rho_0=1$.}
    \label{fig:2}
\end{figure}

Utilizing the Eqs.(\ref{eq:30})-(\ref{eq:32}) we achieve the two metric potential as
\begin{equation}\label{eq:33}
 e^{-\lambda(r)}= \frac{2 (\beta +8 \pi ) \log (r)}{8 \pi -\beta }-C_2 ,
 \end{equation}

\begin{equation}\label{eq:34}
    e^{\nu(r)}= C_3 \left(r(\beta +8 \pi ) \right)^{-\frac{32 \pi }{\beta +8 \pi }}.
\end{equation}
Where $C_2$ and $C_3$ are integrating constants.
Furthermore, by plugging the EoS $p=\rho$ and using Eq.(\ref{eq:34}) into the energy conservation equation (\ref{eq:23}) we have obtained the pressure/matter density for the shell region as,
\begin{equation}\label{eq:35}
    p(r)=\rho(r)= \rho_0 \left(8 \pi  r-\beta  r\right)^{\frac{32 \pi }{8 \pi -\beta }}.
\end{equation}
Where $\rho_0$ is the constant of integration.
Fig.(\ref{fig:2}) shows the variation of pressure or matter density. One can see that the matter density of the shell is monotonically growing up toward the outer boundary of the shell. The shell is made of ultra-relativistic stiff fluid, since the pressure or matter density is monotonically increasing towards the outer surface we can physically interpret that the amount of stiff matter is rising towards the outer border rather than the internal region of the shell.

\subsection{Exterior Region}
The EoS $p=\rho=0$ is believed to be obeyed by the outside of the gravastar, indicating that the external portion of the shell is entirely vacuum. Thus, utilizing Eq.(\ref{eq:16})-Eq.(\ref{eq:17}) along with the Eq.(\ref{eq:19})-Eq.(\ref{eq:20}), we obtain
\begin{equation}\label{eq:36}
    \lambda'+\nu'=0.
\end{equation}
The line element for the outside region may be seen as the well-known Schwarzschild metric, which is provided by the solution to Eq.(\ref{eq:35}) given by,
\begin{equation}\label{eq:37}
    ds^2=\left(1-\frac{2M}{r}\right)dt^2-\left(1-\frac{2M}{r}\right)^{-1} dr^2-r^2 d\Omega^2,
\end{equation}
where $d\Omega^2=(d\theta^2+sin^2\theta d\phi^2)$ and $M$ denotes the total mass of the object.
\subsection{Boundary Condition}

There are two junctions/interfaces in a gravastar configuration. Let us denote the interface between interior space-time and intermediate thin shell (at $r=R_1$) by junction-$I$ and the interface between the intermediate thin shell and exterior space-time (at $r=R_2$) by junction-$II$. It is necessary that the metric functions at these interfaces must be continuous for any stable arrangement. We matched the metric functions at these borders in order to find the unknown constants of our current study like $C_1$, $C_2$, and $C_3$, and we ultimately discovered the values of these constants.

\begin{widetext}
    \begin{itemize}
        \item \textbf{Junction-I :}
        \begin{equation}\label{eq:38}
             \frac{2 (\beta +8 \pi ) \log R_1}{8 \pi -\beta }-C_2 = \frac{2 \rho_c(\beta -4 \pi ) R_1^2  }{3 \alpha }+1,
                \end{equation}
        
        \begin{equation}\label{eq:39}
            C_3 \left(R_1(\beta +8 \pi ) \right)^{-\frac{32 \pi }{\beta +8 \pi }}=C_1 \left[2 \rho_c (4 \pi -\beta ) R_1^2-3 \alpha\right].
        \end{equation}
        \item \textbf{Junction-II :}
        \begin{equation}\label{eq:40}
                 \frac{2 (\beta +8 \pi ) \log R_2}{8 \pi -\beta }-C_2 = 1-\frac{2 M}{R_2},
            \end{equation}
            \begin{equation}\label{eq:41}
                C_3 \left(R_2(\beta +8 \pi ) \right)^{-\frac{32 \pi }{\beta +8 \pi }}=1-\frac{2M}{R_2}.
            \end{equation}
        \item \textbf{Obtained Constants :}
        \begin{equation}\label{eq:42}
        C_3=-\frac{(2 M-R_2) ((\beta +8 \pi ) R_2)^{\frac{32 \pi }{\beta +8 \pi }}}{R_2},
    \end{equation}
    \begin{equation}\label{eq:43}
        C_2= \frac{2 (\beta +8 \pi ) \log (R_1)}{8 \pi -\beta }-\frac{2 \rho_c (\beta -4 \pi )  R_1^2}{3 \alpha }-1,
    \end{equation}
    
    \begin{equation}
        C_1= \frac{(2 M-R_2) ((\beta +8 \pi ) R_1)^{-\frac{32 \pi }{\beta +8 \pi }} ((\beta +8 \pi ) R_2)^{\frac{32 \pi }{\beta +8 \pi }}}{R_2 \left(3 \alpha +2 \beta  \rho_c r^2-8 \pi  \rho_c R_1^2\right)}.
    \end{equation}
     \end{itemize}   
\end{widetext}

Now to find the numerical values of constants $C_1$,$C_2$ and $C_3$ we have considered the astrophysical object PSR J1416-
2230 \cite{P} with $M=1.97M_{\odot}$, internal radius $R_1=10$ and the exterior radius $R_2=10.01$. Apart from that by varying a number of values of model parameter $\alpha$ and $\beta$ we have determined a bunch of numerical values of $C_1$, $C_2$ and $C_3$ which is shown in table-\ref{Tab:1}.\\
With relation to the example of those numerical solutions of constants for some particular parameter choices given above, let's talk about the parameter space of our solution. One could inquire about the following associated problems: 
\begin{enumerate}
    \item For particular choices of $M,R_1,R_2$, will we get a singular free solution always ?
    \item If one varies the model parameter then the results will be unique or not ?
\end{enumerate}

We provide some arguments in answer to these concerns: In the current work, we have selected values for a number of factors to examine the physical behavior of gravastar. It will provide a unique solution for a given value of $M$, $R_1$, and $R_2$, but we have chosen these values in order to satisfy the ratios $\frac{2M}{R_1}<1,\frac{2M}{R_2}<1$ for a stable gravastar model. Besides there are some other criteria like the surface redshift $\mathcal{Z_\mathrm{s}}<2$ and the square of the speed of sound($v_s^2$) must satisfy the inequality $0<v_s^2<1$. Apart from that for avoiding central singularity, we should maintain $\frac{2(\beta-4 \pi) \rho_c r^2 }{3 \alpha } +1 \neq 0$.
Moreover, we have taken $\rho_0=1 $ and $ \rho_c=0.001$ in order to maintain $\rho_0>>\rho_c$ . We are free to choose any $M$, $R_1$, and $R_2$ combination that would provide the same findings as those presented in this research as long as the aforementioned requirements are valid.

\begin{table}\centering
\caption{Different numerical values of constants for  PSR J1416-223 assuming $R_1=10 \,$ km and $R_2=10.01 $\, km.}
\label{Tab:1}

\begin{tabular*}{\columnwidth}{@{\extracolsep{\fill}}rccccccr}
\toprule
$\alpha$  & $\beta$ & $C_1$ & $C_2$ & $C_3$ \\
 \midrule 
  $-4.5$ & $3.4$ & $0.0396871$& $4.91029$ &$2.72477\times10^8$\\
  $-4.6$ & $3.3$ & $ 0.0388762$& $4.86301$ &$2.88649\times10^8$\\
  $-4.7$ & $3.2$ & $0.0380979$& $4.81611$ &$3.05892\times10^8$\\
  $-4.8$ & $3.1$ & $0.0373501$& $4.76958$ &$3.24284\times 10^8$\\
  $-4.9$ & $3.0$ & $0.0366311$& $4.72344$ &$3.4391\times 10^8$\\

\end{tabular*}
\end{table}

\section{Junction Condition and Equation of States}\label{sec:5}
It is established that the gravastar is divided into three regions viz. the interior (I), the intermediate thin shell (II), and the exterior (III). This shell keeps the internal area and outer region connected. So, This region is crucially significant in the construction of the gravastar. According to the fundamental junction requirement, regions I and III must match smoothly at the junction. The derivatives of these metric coefficients may not be continuous at the junction surface, despite the fact that the metric coefficients are continuous there.
In order to calculate the surface stresses at the junction, we will now employ the Darmois-Israel \cite{darmois, Israel, WI} condition. The Lanczos equation  \cite{KL, NS,GP,Mus}  provides the intrinsic surface stress-energy tensor $S_{ij}$ in the following manner:
\begin{align}
    S_{ij}=-\frac{1}{8 \pi}(k_{ij}-\delta_{ij} k_{\gamma \gamma}).
\end{align}
In the above expression, $k_{ij}=K^{+}_{ij}-K^{-}_{ij}$ denotes the discontinuity in some second fundamental expression. Where the second fundamental expression is given by,
\begin{equation}
    K^{\pm}_{ij}=-n^{\pm}_{\sigma} \left(  \frac{\partial x_{\sigma}}{\partial \phi^{i} \partial \phi^{j}} +\Gamma^{l}_{km} \frac{\partial x^l}{\partial \phi^{i}} \frac{\partial x^m}{\partial \phi^{j}} \right),
\end{equation}
where $\phi^{i}$ denotes the intrinsic co-ordinate in the shell area, along with $n^{\pm}$  represents the two-sided unit normal to the surface, which can be written as,
\begin{equation}
    n^{\pm}=\pm \left|g^{lm} \frac{\partial f}{\partial x^{l}} \frac{\partial f}{\partial x^{m}} \right|^{-1/2} \frac{\partial f}{\partial x^{\sigma}},
\end{equation}
with $n^{\gamma} n_{\gamma} =1$.
Utilizing the Lanczos method \cite{KL} the surface energy tensor can be written as $S_{ij}=diag(-\sum, P)$, where the surface energy density and surface pressure are denoted by  $\sum$   and $P$ respectively and are defined by,
 \begin{equation}
     \sum=-\frac{1}{4 \pi \mathcal{R}} \left[\sqrt {e^{-\lambda}}\right]^{+}_{-},
     \label{eq:48}
 \end{equation}
 \begin{equation}
     P=-\frac{\sum}{2}+\frac{1}{16 \pi}\left[\frac{(e^{-\lambda})^{\prime}}{\sqrt {e^{-\lambda}}}\right]^{+}_{-},
     \label{eq:49}
 \end{equation}
 \begin{equation}\label{eq:50}
     \text{Also the EoS}\,(\omega)=\frac{P}{\sum}.
 \end{equation}
Here $-$ and $+$ represent the interior space-time and Schwarzschild space-time respectively. Calculating the Eq. (\ref{eq:48})-(\ref{eq:50}) we get the expression of the above quantities as ,

\begin{widetext}
 \begin{equation}
   \sum=\left(-\frac{1}{4 \pi  R}\right)\left(\sqrt{1-\frac{2 M}{R}}-\sqrt{\frac{2 (\beta -4 \pi ) \text{$\rho $c} R^2}{3 \alpha }+1}\right)
   \label{eq:51}
\end{equation}

\begin{equation}\label{eq:52}
   P=\frac{1}{16 \pi }\left(\frac{2 M}{R^2 \sqrt{1-\frac{2 M}{R}}}-\frac{4 (\beta -4 \pi ) \rho_c R}{3 \alpha  \sqrt{\frac{2 (\beta -4 \pi ) \rho_c R^2}{3 \alpha }+1}}\right)-\frac{1}{2} \left(-\frac{1}{4 \pi  R}\right)\left(\sqrt{1-\frac{2 M}{R}}-\sqrt{\frac{2 (\beta -4 \pi ) \rho_c R^2}{3 \alpha }+1}\right),
\end{equation}

\begin{equation}
    \omega=\frac{\frac{1}{16 \pi }\left(\frac{2 M}{R^2 \sqrt{1-\frac{2 M}{R}}}+\frac{4 (4 \pi -\beta ) \rho_c R}{\alpha  \sqrt{\frac{6 (\beta -4 \pi ) \rho_c R^2}{\alpha }+9}}\right)}{\left(-\frac{1}{4 \pi  R}\right)\left(\sqrt{1-\frac{2 M}{R}}-\sqrt{\frac{2 (\beta -4 \pi ) \rho_c R^2}{3 \alpha }+1}\right)}-\frac{1}{2}.
\end{equation}
\end{widetext}
There is some set of conditions known as energy conditions that must be applied in order for a geometric structure to be physically viable.
The well-recognized energy criterion are :
\begin{enumerate}
    \item \textbf{NEC:} $\sum+P> 0$,
    \item \textbf{WEC:} $\sum> 0$,\,$\sum+P>0$ ,
    \item \textbf{SEC:} $\sum+P> 0$,$\sum+3P> 0$,
    \item \textbf{DEC:}$\sum > 0,\sum \pm P> 0$ .
\end{enumerate}
    
\begin{figure}
\includegraphics[scale=0.43]{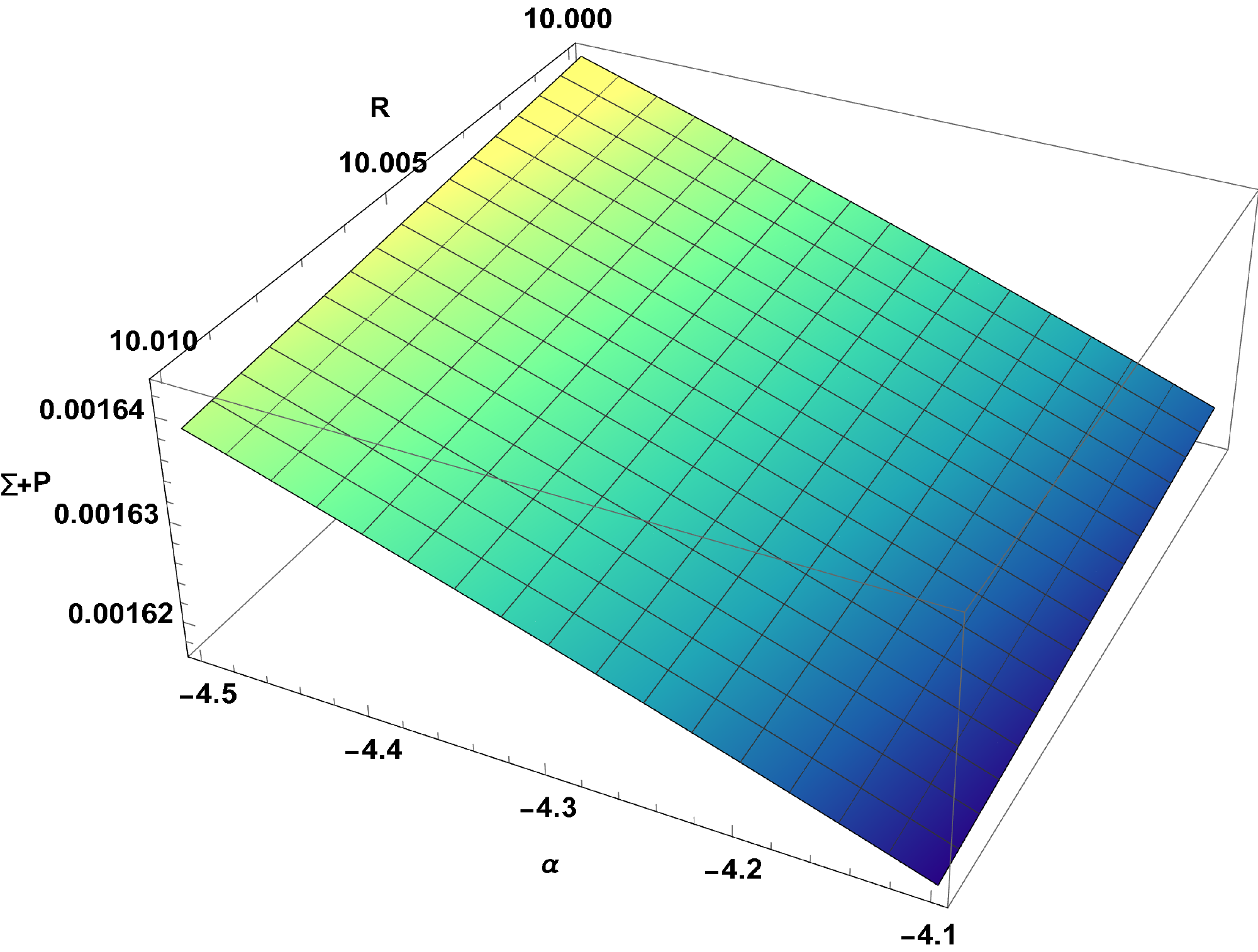}
    \caption{Evolution of the NEC ($\sum+P$)  by varying model parameter $\alpha$ .}
    \label{fig:3}
\end{figure}

The presented model is physically feasible if these energy criteria are established. Here, we're investigating to see if the null energy requirement, which guarantees the presence of ordinary or exotic matter in the thin shell, is satisfied or not. In this context, it is noteworthy to say that violation of null energy conditions (NEC) leads to violation of other energy conditions. It is illustrated in Fig.(\ref{fig:3}) that the NEC is satisfied over a range of model parameter values throughout the entire region.

\begin{figure}
\includegraphics[scale=0.43]{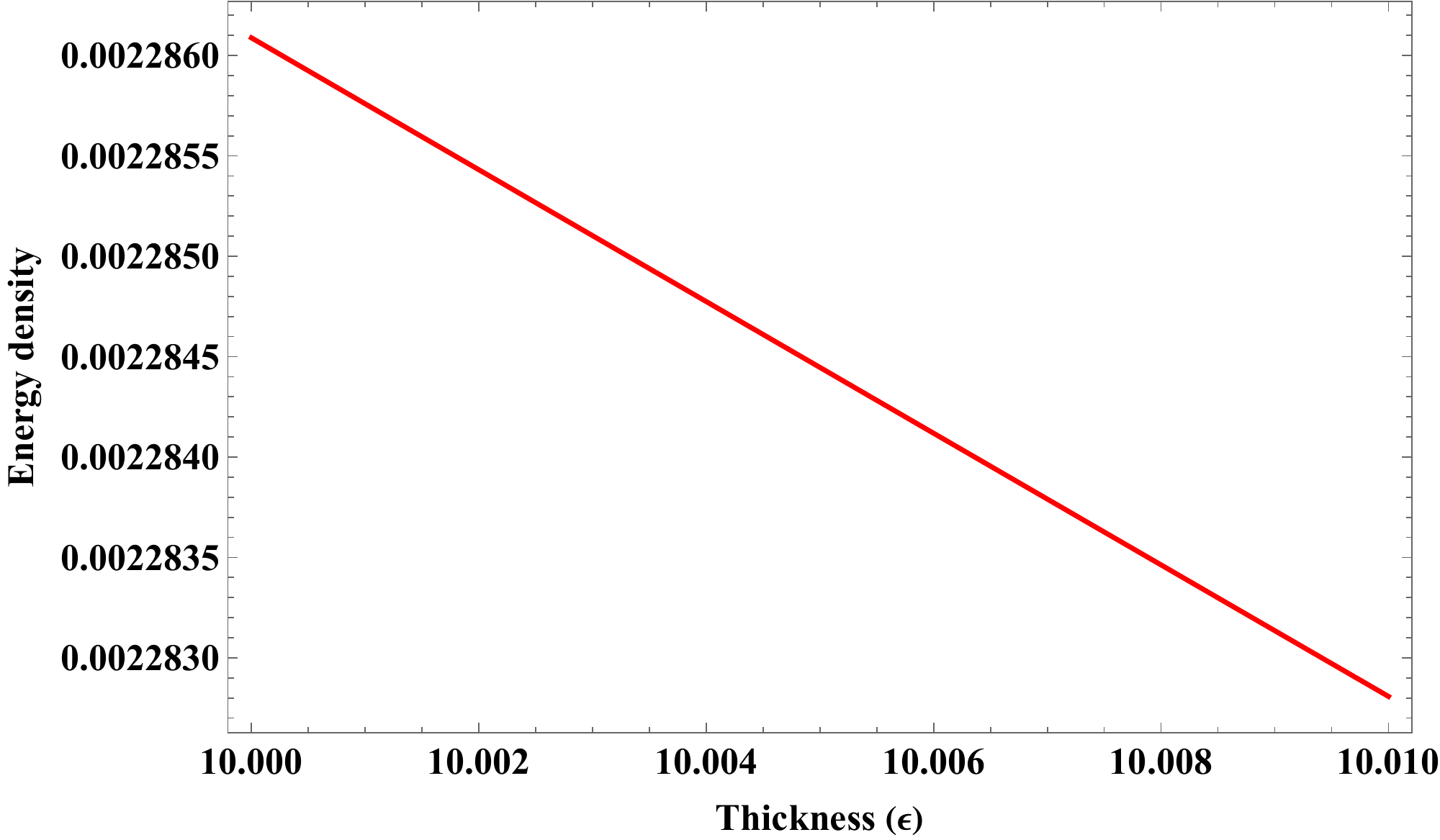}
    \caption{ variation of the Surface energy density ($\sum$)  with respect to thickness $\epsilon(\text{in km})$ for $\alpha=-4.5, \beta=3.4$.}
    \label{fig:4}
\end{figure}

 Besides that, we have plotted the variation of surface energy density with respect to the thickness parameter($\epsilon$) which shows that the surface energy density is monotonically decreasing towards the boundary of the shell.
The mass of the thin shell now is easily determined using the equation for the surface energy density given by,

\begin{multline}\label{eq:54}
    m_s= 4\pi R^2 \sum = -R\left(\sqrt{1-\frac{2 M}{R}}-\sqrt{\frac{2 (\beta -4 \pi ) \rho_c R^2}{3 \alpha }+1}\right).
\end{multline}

Now for determining the real value of shell mass, we have the inequality, $m_s>0$ from which we get the upper bound of the radius as $R<\left(\frac{3 M \alpha}{\rho_c (4 \pi- \beta)}\right)^\frac{1}{3}.$ Thus we get the limiting value on the radius as
\begin{equation}\label{eq:55}
    2M<R<\left(\frac{3 M \alpha}{\rho_c (4 \pi- \beta)}\right)^\frac{1}{3}.
\end{equation}

\section{Physical Features of the Model}\label{sec:6}

\subsection{Proper thickness}
\begin{figure}
\includegraphics[scale=0.41]{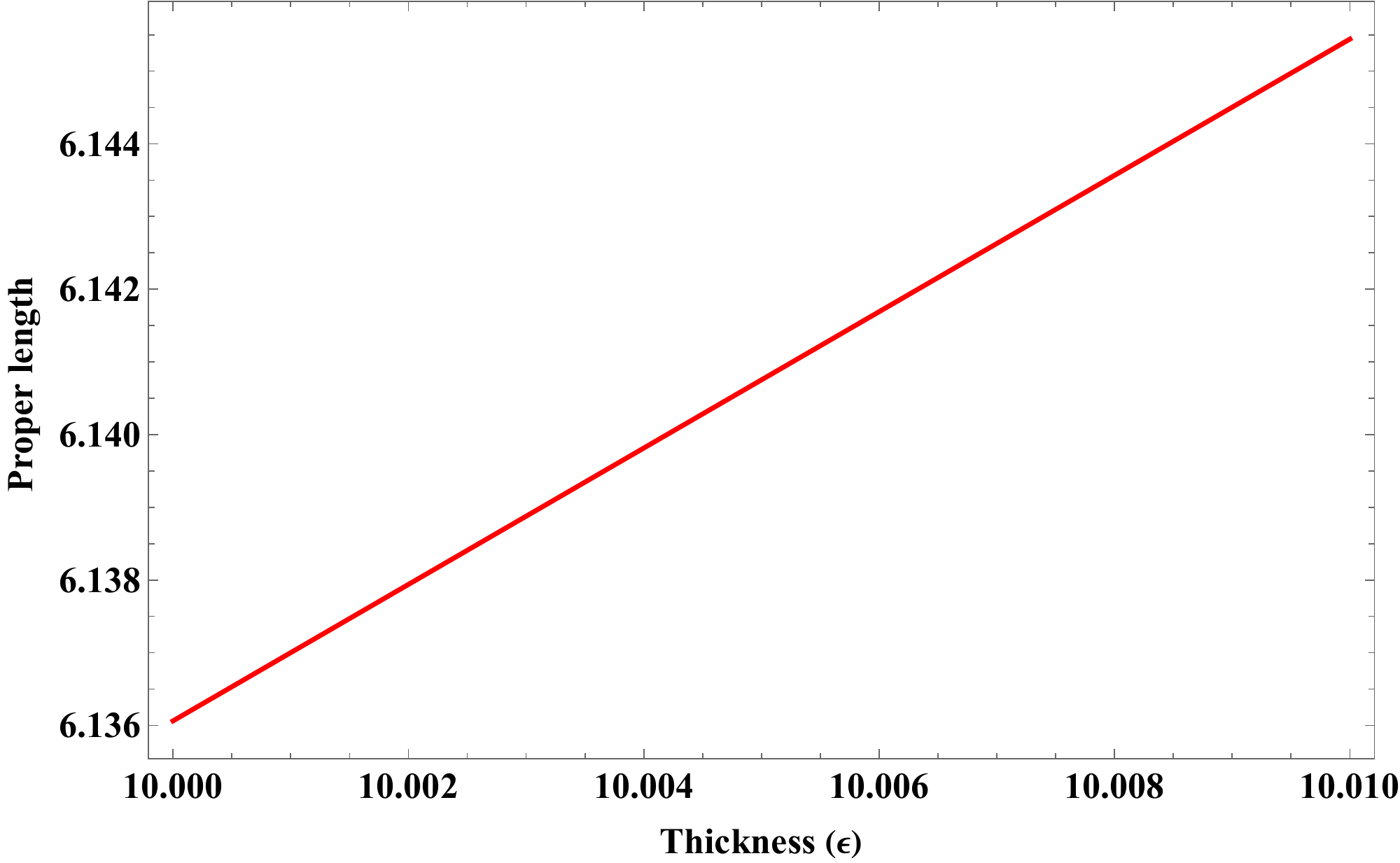}
    \caption{ variation of the proper length ($l$)  with respect to thickness $\epsilon(\text{in km})$ for $\alpha=-4.5$ and $\beta=3.4$.}
    \label{fig:5}
\end{figure}

According to Mazur and Mottola's hypotheses \cite{mazur,mottola}, the stiff fluid of the shell is positioned between the meeting of two space-times. The length of the shell ranges from $R_1=R$ (which is the phase barrier between the interior area and intermediate thin shell) up to $R_2=R+\epsilon$ (which is the phase border between the exterior space time and intermediate thin shell). So, using the following formula, one can find the required length or proper thickness of the shell as well as the proper thickness between these two interfaces:
\begin{equation}
   \begin{gathered}\label{eq:56}
      \textit{l}=\int^{R+\epsilon}_{R} \sqrt{e^{\lambda}}dr, \\
    =\int^{R+\epsilon}_{R} \sqrt{\frac{\beta -8 \pi }{(8 \pi -\beta ) C_2-2 (\beta +8 \pi ) \log (r)}} \, dr,\\ 
    =\left[-e^{-\frac{(\beta -8 \pi ) C_2}{2 (\beta +8 \pi )}}\sqrt{\frac{\pi  (\beta -8 \pi )}{2 (\beta +8 \pi )}}  \text{Erf}\sqrt{\frac{(8 \pi -\beta ) C_2}{2 (\beta +8 \pi )}-\log (r)}
\right]_{R}^{R+\epsilon}
   \end{gathered}
\end{equation}
The variation of the proper length with respect to the thickness parameter $\epsilon$ is given in Fig.(\ref{fig:5}). The figure demonstrates that the proper length rises monotonically as shell thickness increases.

\subsection{Energy}
 The energy of the shell can be calculated by the formula,
\begin{equation}
   \begin{gathered}
       E=\int^{R+\epsilon}_R 4\pi r^2 \rho \, dr, \\ E=\int^{R+\epsilon}_R 4\pi r^2 \rho_0 \left(8 \pi  r-\beta  r\right)^{\frac{32 \pi }{8 \pi -\beta }}  \,\, dr,
       \\ 
     E  =\frac{4 \pi  \rho_0 r^2 ((8 \pi -\beta ) r)^{\frac{32 \pi }{8 \pi -\beta }+1}}{56 \pi -3 \beta }.
       \label{eq:59}
   \end{gathered}
\end{equation}

\begin{figure}
\includegraphics[scale=0.41]{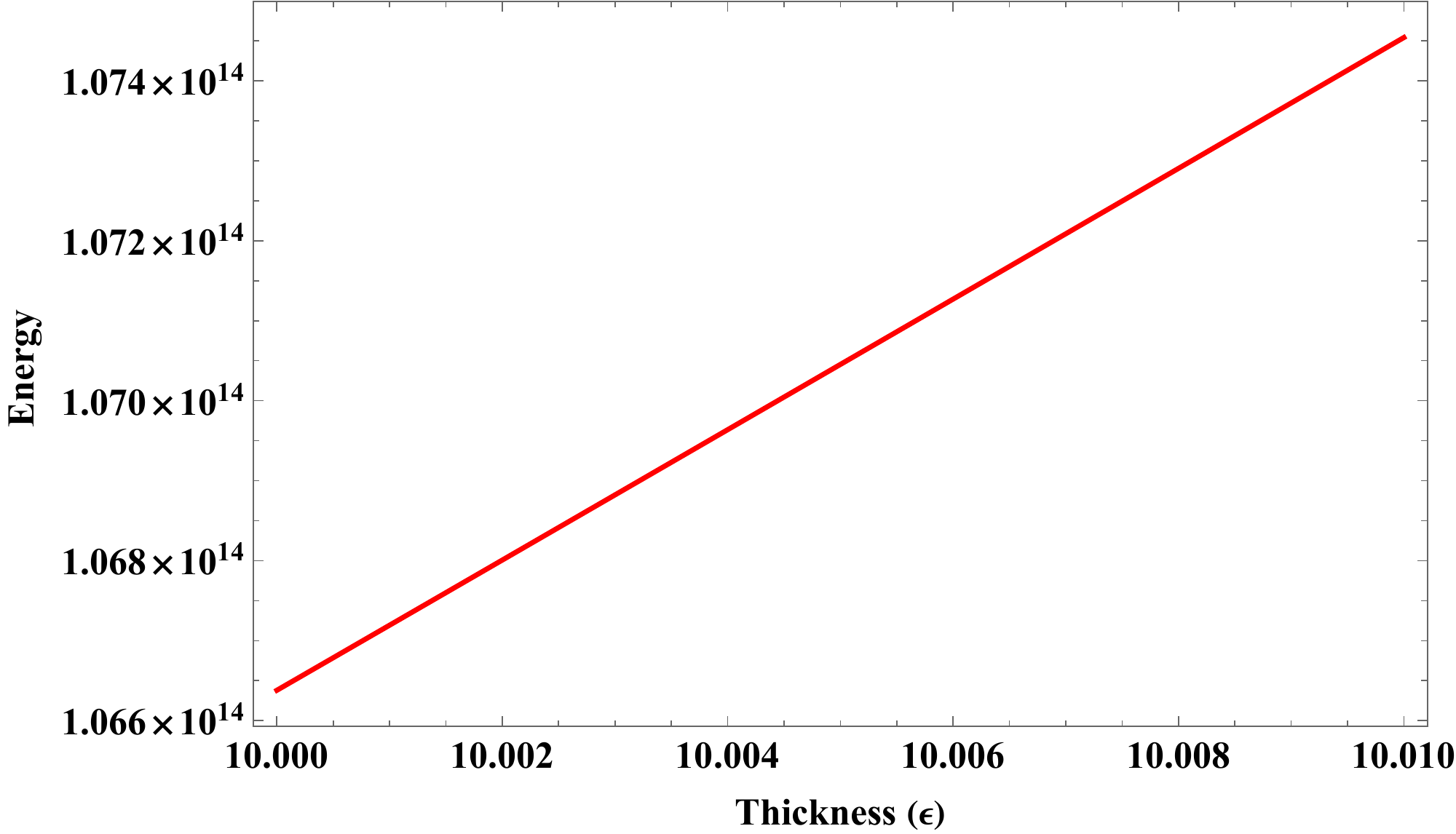}
    \caption{variation of energy $(E)$ with respect to thickness$\epsilon(\text{in km})$ for $\alpha=-4.5$ and $\beta=3.4$.}.
    \label{fig:6}
\end{figure}
The variation of the shell energy is illustrated in Fig.(\ref{fig:6}). In this graph, it can be observed that the energy rises as the shell's thickness increases. The fluctuation of energy is comparable to the fluctuation in matter density. It meets the requirement that the energy of the shell must be increased as the radial distance increases.

\subsection{Entropy}

The stable configuration for a single condensate area is zero entropy density, which is present in the gravastar's innermost region. Entropy on the intermediate thin shell can be calculated using the formula according to Mazur and Mottola's work \cite{mazur,mottola},

\begin{equation}
   \mathcal{S}= \int^{R+\epsilon}_{R} 4 \pi r^2 s(r) \sqrt{e^{\lambda}} dr,
\end{equation}

here the entropy density at local temperature $T(R)$ is given by the expression $s(r)=\frac{\gamma^2 K_{B}^2  T(R)}{4 \pi \hbar^2 }= \gamma \sqrt{p/{2\pi}}$ where $\gamma$ is the dimensionless parameter. In this work, we have considered geometrical units i.e. $G=c=1$ as well as Planckian units $K_{B}=1, \hbar =1$.
Our estimates of the entropy of the thin shell are limited to the second-order term of the thickness parameter, i.e. the order $\epsilon^2$, using Taylor series approximation. Ultimately, we have calculated the intermediate thin shell's entropy as follows:

\begin{widetext}
    \begin{equation}
        \begin{gathered}
           \mathcal{S}= 2 \sqrt{2 \pi } \gamma  r^2 \epsilon  \sqrt{\frac{(\beta -8 \pi ) \rho_0 (8 \pi  r-\beta  r)^{\frac{32 \pi }{8 \pi -\beta }}}{(8 \pi -\beta ) C_2-2 (\beta +8 \pi ) \log (r)}}\,\,+\\ \frac{\epsilon ^2 \left(2 \sqrt{2 \pi } \gamma  r ((8 \pi -\beta ) (\beta -2 \beta  C_2+8 \pi  (4 C_2+1))-4 (16 \pi -\beta ) (\beta +8 \pi ) \log (r)) \sqrt{\frac{(\beta -8 \pi ) \rho_0 ((8 \pi -\beta ) r)^{-\frac{32 \pi }{\beta -8 \pi }}}{(8 \pi -\beta ) C_2-2 (\beta +8 \pi ) \log (r)}}\right)}{2 \left((\beta -8 \pi )^2 C_2+2 \left(\beta ^2-64 \pi ^2\right) \log (r)\right)}.
        \end{gathered}
    \end{equation}
\end{widetext}

\begin{figure}
\includegraphics[scale=0.41]{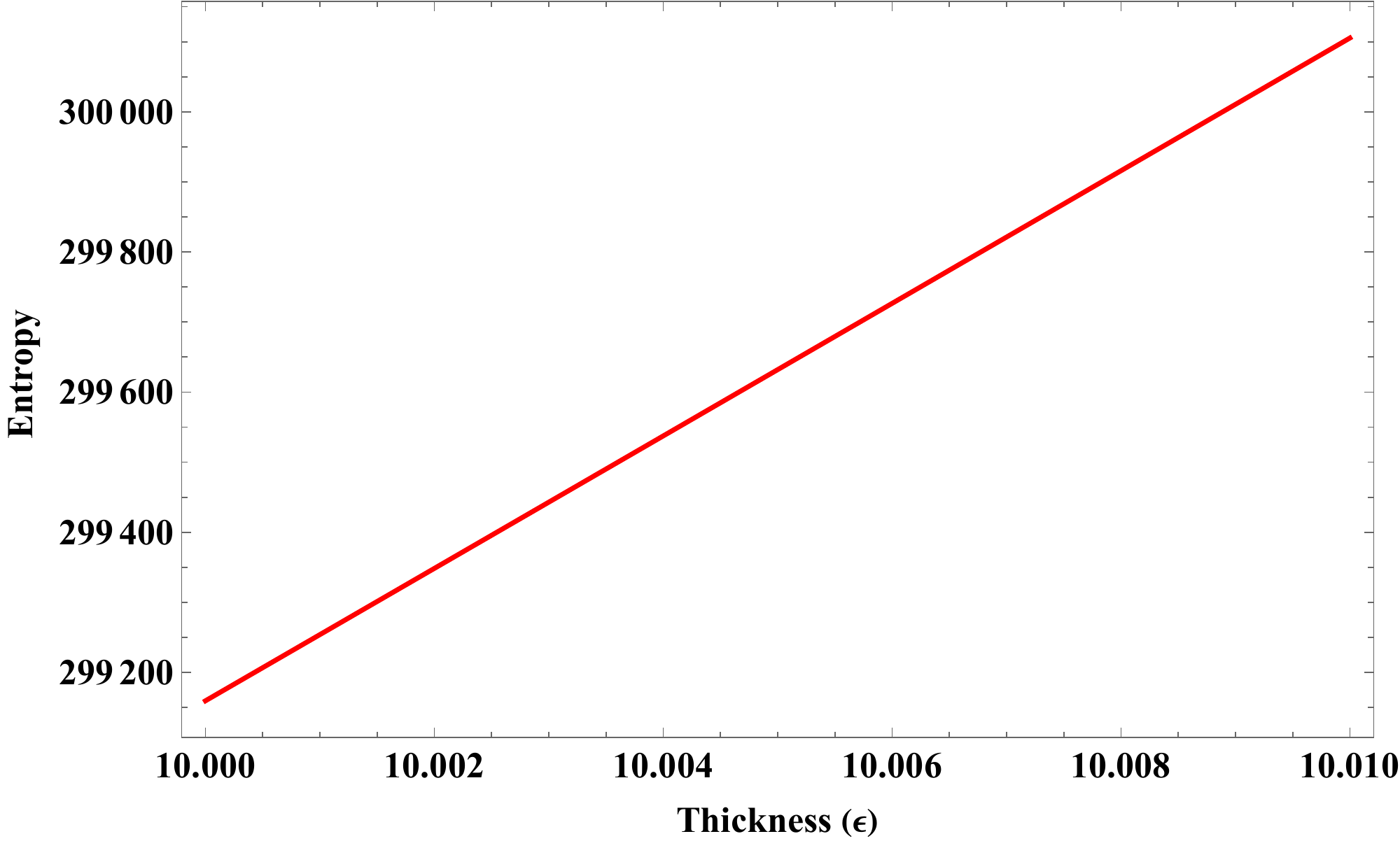}
    \caption{ variation of the entropy ($\mathcal{S}$)  with respect to thickness $\epsilon(\text{in km})$ for $\beta=3.4$, $\alpha=-4.5$.}
    \label{fig:7}
\end{figure}

Fig.(\ref{fig:7}) depicts the evolution of the shell entropy, which shows the growing behavior of the shell entropy with regard to thickness ($\epsilon$). Another acceptable condition is that entropy should reach its greatest value on the surface for a stable gravastar configuration which is demonstrated in our analysis.

\section{Stability of the stellar model}\label{sec:7}
In this section, we are interested to investigate the stability of the thin shell gravastar model by analyzing some physical parameters.

\subsection{Study of Herrera's cracking concept}
Recent observational data appear to indicate that the cosmos is expanding more quickly than before\cite{AR, SJ, NA}. If general relativity is taken to be the right theory of gravity characterizing the behavior of the universe on a large scale, then the energy density and pressure of the cosmos should violate the strong energy condition. The stable or unstable configuration of gravastars could be analyzed through the nature of $\eta$, where $\eta$ is an effective parameter that can be interpreted as the square of the speed of sound i.e. $\eta=v_s^2$ \cite{sharif, SoS}.
For a stable system $\eta$ should satisfy $0<\eta\leq 1$. The speed of sound shouldn't be higher than the speed of light, as is clear. However, this restriction might not be met on the surface layer to test the gravastar's stability. The square of the speed of sound is defined by,
\begin{equation}
    \eta=v_s^2=\frac{P^{\prime}}{\sum'}.
\end{equation}
Where ' represents the derivative w.r.t the radial coordinate. As a result, by using (\ref{eq:51},\ref{eq:52}) we examine the parameter's sign to determine the stability of gravastar configurations. We utilize the graphical behavior since the mathematical expression of $\eta$ is complicated. 

\begin{figure}
\includegraphics[scale=0.55]{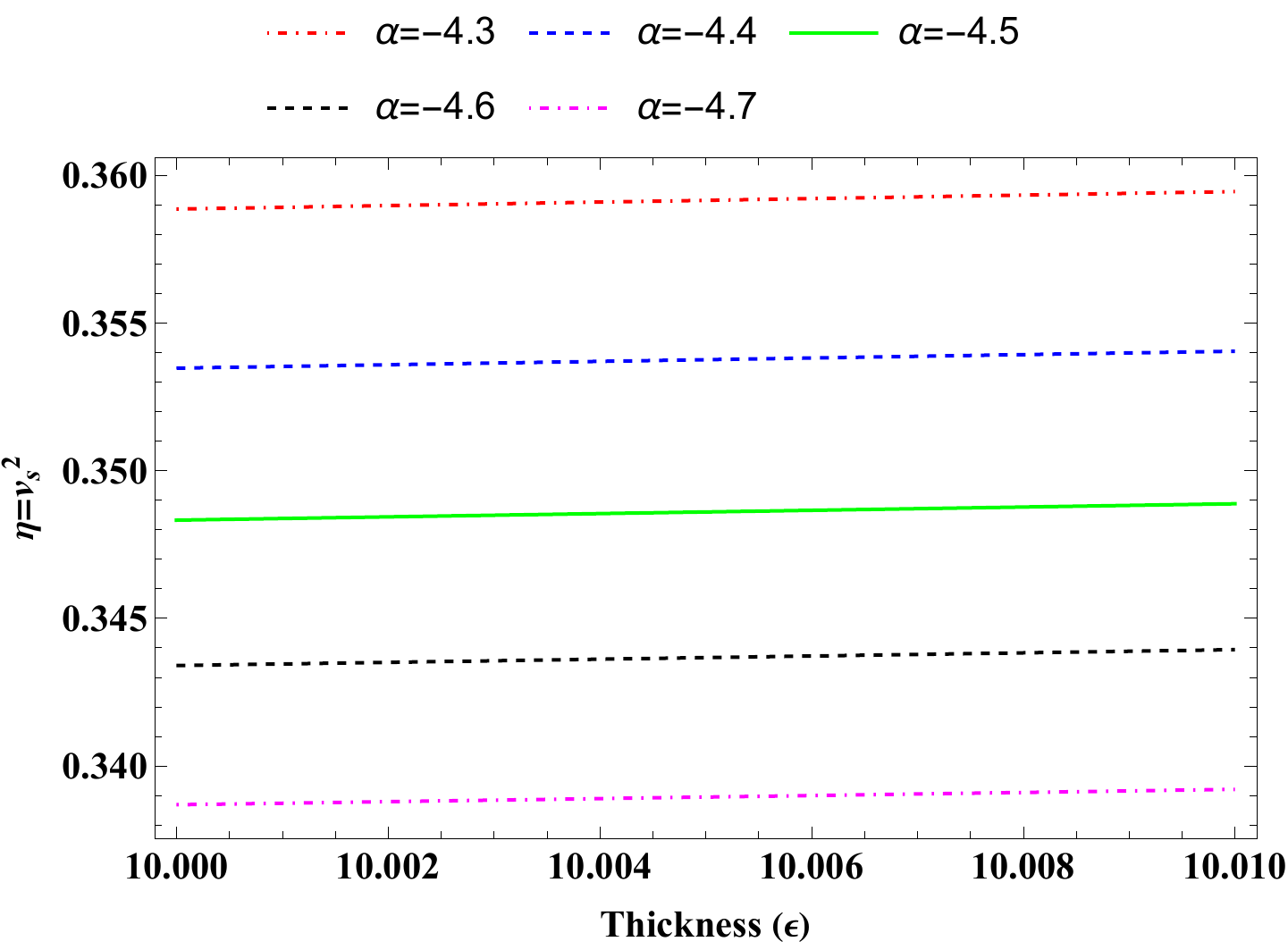}
    \caption{ variation of $\eta$  with respect to thickness $\epsilon(\text{in km})$ for different values of $\alpha$}
    \label{fig:8}
\end{figure}
From Fig.(\ref{fig:8}) it could be noticed that the effective parameter $\eta$ satisfies the inequality $0<\eta\leq 1$ throughout the entire shell region. Here we have varied the model parameter $\alpha$ and we observed for each value of $\alpha$ our model behaves physically stable. One more important observation to mention is that whenever the value of $\alpha$ increases the parameter $\eta \to 1$. So as the model parameter value rises our proposed gravastar model approaches the unstable situation. 

\subsection{Surface Redshift}

The study of a gravastar's surface redshift is one of the most basic ways to understand the stability and detection of the object. The formula $Z_s=\frac{\Delta \lambda}{\lambda_e}=\frac{\lambda_0}{\lambda_e}$ could be used for determining the gravitational surface redshift of the gravastar, where $\lambda_0$ and $\lambda_e$ represents the wavelength detected by the observer and emitted from the source.
Buchdahl \cite{HA, N} proposed that the value of surface redshift should not be more than 2 for an isotropic, stable, perfect fluid distribution.
However, Ivanov \cite{ivanov} claimed that for anisotropic fluid dispersion, it might go as high as $3.84$. Other than that Barraco and Hamity \cite{hamity} showed that for an isotropic fluid distribution, $Z_s\leq 2$ holds when the cosmological constant is absent. Bohmer and Harko \cite{bohmer} though, showed that in the presence of the anisotropic star's cosmological constant, $Z_s \leq 5$. Now in our case, we have obtained the surface redshift by the following formula,

\begin{equation}
    Z_s=-1+\frac{1}{\sqrt{g_{tt}}}=\frac{1}{\sqrt{C_3 ((\beta +8 \pi ) r)^{-\frac{32 \pi }{\beta +8 \pi }}}}-1 .
\end{equation}

\begin{figure}
\includegraphics[scale=0.45]{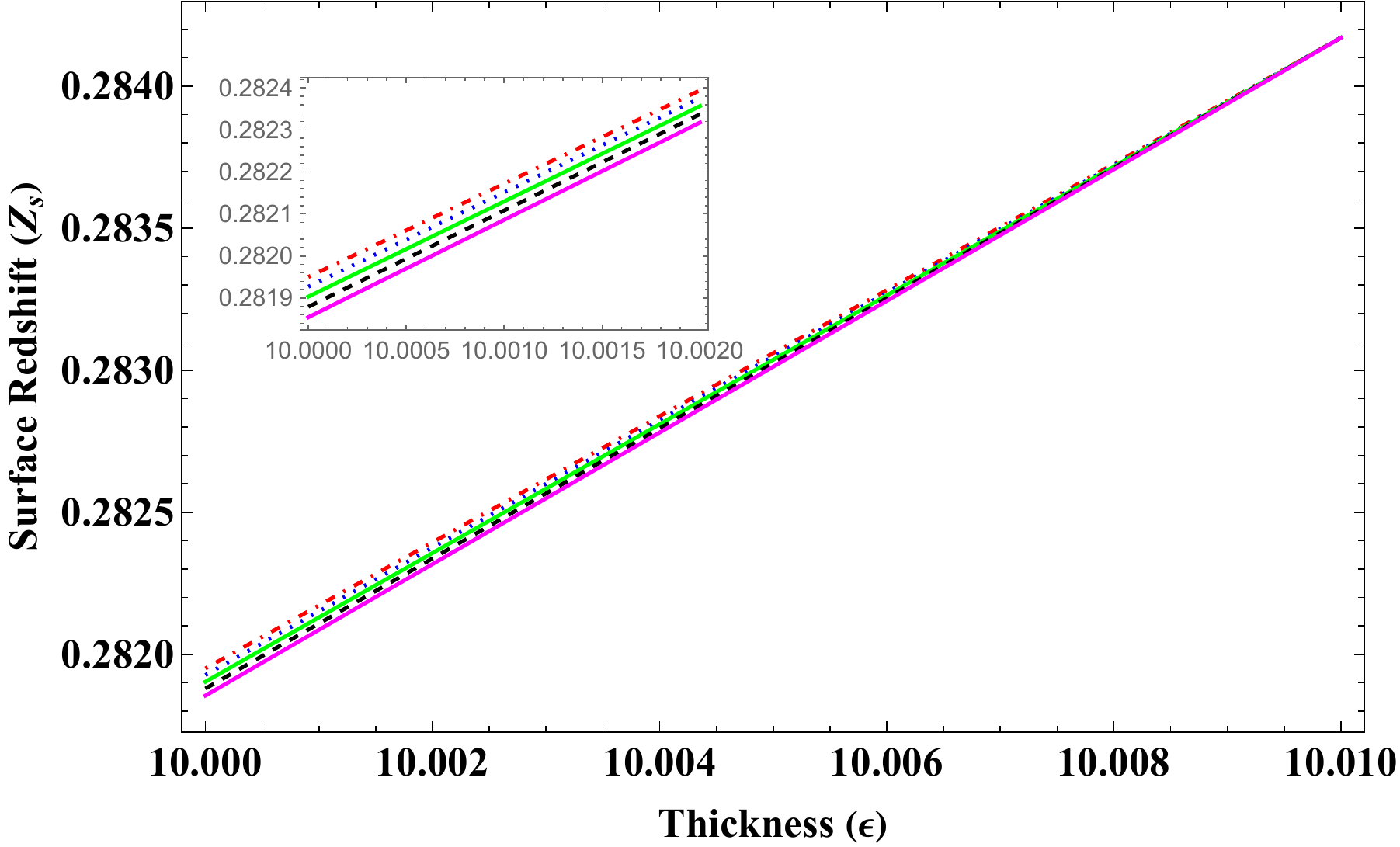}
    \caption{ variation of $Z_s$  with respect to thickness $\epsilon(\text{in km})$ for different values of model parameter $\alpha \,\text{and} \, \beta$. Red Dotdashed ($\alpha=-4.1, \beta=3.9$), Blue Dotted ($\alpha=-4.4, \beta=3.6$), Green Tan ($\alpha=-4.7, \beta=3.3$), Black Dashed ($\alpha=-5.0, \beta=3.0$), Magenta Thickness  ($\alpha=-5.3, \beta=2.7$)}
    \label{fig:9}
\end{figure}

The graphical analysis of $Z_s$ is given in Fig.(\ref{fig:9}). We have varied the model parameter $\alpha$ and $\beta$ for analyzing the maximum possibility case of $Z_s$ and in each case, it is noticed that $Z_s<1$. Consequently, we can thus assert that the current gravastar model is both physically stable and appropriate in the $f(Q, T)$ framework.

\subsection{Entropy Maximization:}
Each quasi-black hole (QBH) candidate must be stable to constitute a physically feasible endpoint of gravitational collapse \cite{MazurPO}. Now, in order to verify the stability of the current investigation of gravastar in $f(Q,T)$ gravity, we have used the entropy maximization method recommended by Mazur and Mottola \cite{mazur,mottola}. Since the shell region is only the non vacuum region with stiff fluid and contains the positive heat capacity so the solution should thermodynamically stable here. To check the stability we will use the entropy maximization technique in the shell region. For maximizing the entropy function at first the first variation of the entropy function should vanish at the boundaries of the shell i.e. $\delta \mathcal{S}=0$ at $r=R_1$ and $r=R_2$. After that we have checked the nature of the second derivative i.e. of $\delta^2 \mathcal{S}$ by it's sign for all the variation of $M(r)$. The entropy function is given by,

\begin{equation}
    \mathcal{S}=\frac{\gamma k_B}{ \hbar G} \int_{R_1}^{R_2} r d r\left(2\frac{ d M}{d r}\right)^{\frac{1}{2}} \frac{1}{\sqrt{1-\frac{2 M(r)}{r}}}.
\end{equation}
A necessary and sufficient condition for the dynamic stability of a static, spherically symmetric solution of the field problem is thermodynamic stability in the context of a hydrodynamic treatment.
The second derivative of the entropy function is given by,
\begin{equation}
    \begin{gathered}\label{eq:63}
        \delta^2 \mathcal{S}=\frac{\gamma k_B}{\hbar G} \int_{R_1}^{R_2} r d r\left(2 \frac{d M}{d r}\right)^{-\frac{3}{2}} \left(1-\frac{2 M}{r}\right)^{-\frac{1}{2}}\\ \left\{-\left[\frac{d(\delta M)}{d r}\right]^2+2 \frac{(\delta M)^2}{r^2 \left(1-\frac{2 M}{r}\right)^2} \frac{d M}{d r}\left(1+2 \frac{d M}{d r}\right)\right\}.
    \end{gathered}
\end{equation}

With the help of equation \ref{eq:33}, \ref{eq:37} and \ref{eq:43} we have determined the functional value of $M(r)$ as,

\begin{equation}
    \begin{gathered}
        M(r)=\frac{r \left(3 \alpha  (\beta +8 \pi ) \ln (\frac{R_1}{r})+\left(\beta ^2-12 \pi  \beta +32 \pi ^2\right) \rho_c R_1^2\right)}{3 \alpha  (8 \pi -\beta )}.
    \end{gathered}
\end{equation}
Now if we consider the linear combination of $M(r)$ as $\delta M=\chi_0 \psi$ where $\psi$ becomes vanish at the boundaries $R_1$ and $R_2$, then integrating Eq.(\ref{eq:63}) partially by using the diminishing of the variation $\delta M$ we get,

\begin{equation}\label{eq:65}
    \delta^2 \mathcal{S}=-\frac{\gamma k_B}{\hbar G} \int_{R_1}^{R_2} \frac{r d r} {\sqrt{\left(1-\frac{2 M}{r}\right)}}\left(2 \frac{d m}{d r}\right)^{-\frac{3}{2}} \chi_0^2\left(\frac{d \psi}{d r}\right)^2<0 .
\end{equation}

It is evident from the above expression that for any radial variations that vanish at the endpoints of the shell's boundaries, the entropy function in $f(Q, T)$ gravity reaches its maximum value. We may thus draw the conclusion that a perturbation in the gravastar's intermediate shell area's fluid leads to a decrease in entropy in region II, which indicates the idea that our solutions are stable against minor perturbations with the specified endpoints. In essence, the stability of the gravastar is unaffected by the effect of $f(Q,T)$ gravity.

\section{Discussion and Conclusion}\label{sec:8}
Following the model put forward by Mazur-Mottola \cite{mazur,mottola} within the context of general relativity, we have developed a unique stellar model of a gravastar under the theory of $f(Q, T)$ gravity in this research. There are three different regions namely interior region, intermediate thin shell, and exterior space-time with three different EoS. The interior region fully consists of dark energy as hypothesized by \cite{mazur,mottola}. The following are some of the gravastars' crucial characteristics:
\smallskip

\begin{itemize}
    \item \textbf{Interior Region :}
    Using the EoS (\ref{eq:25}) we have derived two non-singular metric potentials (\ref{eq:27},\ref{eq:28}) from the described field equation in $f(Q,T)$ gravity. The metric potentials are finite and remain positive throughout the entire interior region. This confirms our proposed gravastar model in $f(Q,T)$ gravity is able to  devoid of the concept of central singularity in CBH.
    \item\textbf{Intermediate thin shell:}
    We have estimated the metric potentials in the region of the shell by using the thin shell approximation. Eq. (\ref{eq:33}) and Eq.(\ref{eq:34}) indicate that two metric potentials remain finite as well as positive throughout the entire shell. 
        \begin{itemize}
            \item \textbf{Pressure or Matter density:} Apart from that using the energy conservation equation (\ref{eq:23}), we have derived the pressure or matter density (\ref{eq:35}) in the shell. Fig.(\ref{fig:2}) represents the variation of the pressure or matter density with respect to the thickness parameter ($\epsilon$). One can see that the matter density of the shell is monotonically growing up toward the outer boundary of the shell. The shell is made of ultra-relativistic stiff fluid, since the pressure or matter density is monotonically increasing towards the outer surface we can physically interpret that the amount of stiff matter is rising towards the outer border rather than the internal region of the shell. That is why the shell's outer boundary becomes denser than the interior border.
        \end{itemize}
    
    \item\textbf{Junction Condition and EoS :}
    The junction requirement for the formation of a thin shell is taken into account between the interior and external space-times. We analyze the variation in surface energy density  with respect to the thickness parameter ($\epsilon$) using the Darmois-Israel junction condition, as shown in Fig.(\ref{fig:4}). The surface energy density increases towards the outer boundary of the shell. Besides that, in Fig.(\ref{fig:3}), we have verified that the NEC is satisfied over a range of model parameter values throughout the entire shell. It confirms the  presence of ordinary or exotic matter in the shell. Apart from that, we get the limiting value of radius (\ref{eq:55}) using the concept of determining the real value of shell mass.
    \item  \textbf{Physical Features of the Model:}  Using the geometrical quantity of the intermediate thin shell we have analyzed some physical properties of the thin shell.
        \begin{itemize}
            \item \textbf{Proper length:}
            The variation of the proper length with respect to the thickness parameter $\epsilon$ is given in Fig.(\ref{fig:5}) and in Eq.(\ref{eq:56}). The figure demonstrates that the appropriate length rises monotonically as shell thickness increases. This monotonically increasing behavior of proper length of gravastar is similar to the work which has been done in modified gravity \cite{Ghosh,ghosh}. 
            \item \textbf{Energy :}
            The variation of the shell energy is illustrated in Fig.(\ref{fig:6}). In this graph, it can be observed that the energy rises as the shell's thickness increases. The fluctuation of energy is comparable to the fluctuation in matter density. It meets the requirement that the energy of the shell increases as the radial distance increases. 
            \item \textbf{Entropy :}
            Fig.(\ref{fig:7}) depicts the evolution of the shell entropy, which shows the growing behavior of the shell entropy with regard to thickness ($\epsilon$). Another acceptable condition is that entropy should reach its greatest value on the surface for a stable gravastar configuration which is demonstrated in our analysis. For comparison of the energy and entropy of the gravastar model with the previous work \cite{amit} one can see that the energy and entropy should reach their maximum value at the boundary of the shell which is established in our study.
        \end{itemize}
    \item \textbf{Stability of stellar model:} Finally we have verified the stability of our proposed stellar model through the study of the Herreras cracking concept and by the study of the surface redshift analysis method. After that we have used the entropy maximization technique to determine the stability of the gravastar.
        \begin{itemize}
            \item \textbf{Herrera's cracking concept:}
            We have analyzed the stability of gravastar by the nature of the effective parameter $\eta$. In Fig.(\ref{fig:8}) it is clear that for each value of $\alpha$ the square of the speed of the sound remains positive and not exceeding 1. Moreover, we can see that for rising the value of the $\alpha$ parameter the model approaches instability.
            \item \textbf{Surface Redshift:}
            Lastly, we used surface redshift analysis to check the stability of our recently suggested model. The surface redshift ($Z_s$) for any physically stable star arrangement should always be smaller than 2. By varying the model parameter $\beta$ we have plotted the surface redshift with respect to the thickness parameter ($\epsilon$) which is given in Fig.(\ref{fig:9}) and in each case $Z_s<1$. It demonstrates that our suggested model is stable under $f(Q, T)$ gravity.
        \end{itemize}
        \begin{itemize}
            \item \textbf{Entropy Maximization :} Here we have applied the entropy maximization technique for checking the stability of the gravastar system. For maximizing the entropy function at first  the first variation of the entropy function is set to be zero at the boundaries of the shell i.e. $\delta \mathcal{S}=0$ at $r=R_1$ and $r=R_2$. After that, we checked the nature of the second derivative i.e. of $\delta^2 \mathcal{S}$ by its sign for all the variations of $M(r)$. Eq.\ref{eq:65} takes a negative value which represents that the entropy attains its maximum value for all variations of the radial parameter. This further indicates the stability of our gravastar model in $f(Q, T)$ gravity. One can check the stability of the gravastar model for the entropy maximization technique in \cite{mazur,amit}.
        \end{itemize}
\end{itemize}

We can draw the conclusion that the gravastar might exist within the constraints of $f(Q, T)$ gravity. In comparison to past work on gravastars, we have extended the thin shell approximation up to the second order, which provides a more accurate analytical solution for determining the physical parameters of the shell. As well as we have applied a new technique  of Herrera's cracking concept to check the stability of our proposed model in $f(Q,T)$ gravity. We may conclude that the $f(Q, T)$ theory of gravity was effectively used in the current study on the gravastar by making this finding. The problem of the black hole's event horizon and the central singularity is promptly solved by a group of physically plausible, non-singular gravastar solutions.

\smallskip

\textbf{Data availability} There are no new data associated with this article.

\acknowledgments 
SP \& PKS  acknowledges the National Board for Higher Mathematics (NBHM) under the Department of Atomic Energy (DAE), Govt. of India for financial support to carry out the Research project No.: 02011/3/2022 NBHM(R.P.)/R \& D II/2152 Dt.14.02.2022. PKS thanks Transilvania University of Brasov for Transilvania Fellowship for Visiting Professors. We are very grateful to the honorable referees and the editor for the illuminating suggestions that have significantly improved our research quality and presentation.

\end{document}